\documentstyle[12pt]{article}
\def\Journal#1#2#3#4{{#1} {\bf #2}, #3 (#4)}

\def\NPB{{\em Nucl. Phys.} B}
\def\PLB{{\em Phys. Lett.}  B}
\def\PRL{\em Phys. Rev. Lett.}

\begin{document}

\begin{flushright}

ITEP/TH-22/99

\end{flushright}
\vspace*{0.5cm}                        
\begin{center}{\bf  AN INFINITE SET OF TREE AMPLITUDES IN HIGGS-YANG-MILLS}\\
\vspace*{0.5cm}
{\bf K.G. SELIVANOV\\
ITEP, B.Cheremushkinskaya,25, Moscow, Russia }
\end{center}

\abstract{It is pointed out that the Parke-Taylor or maximally 
helicity violating amplitudes in the pure Yang-Mills can, after some
specifications, be interpreted as amplitudes of scattering of massive vector
bosons in the Higgs-Yang-Mills system.}

\vspace*{2cm}

Although, in general, multi-particle tree amplitudes are complicated things
(because the number of various Feynman diagrams grows very fast with the 
number of legs), see, e.g., the review \cite{mapa} and more recent reviews
\cite{dixon}, \cite{bern2}, in some instances one can explicitly 
construct infinite sets of tree amplitudes, as it was done in \cite{PT}, 
\cite{B}, \cite{KS4}.

The basic example is  \cite{PT}, where nice explicit expressions for the 
so-called maximally helicity violating (MHV) amplitudes in Yang-Mills (YM)
have been conjectured (these expressions have been proved in \cite{BG}).
MHV amplitudes are gluonic amplitudes with arbitrary number of legs
and most of the external gluons in the same helicity states. More precisely,
MHV amplitudes describe scattering of two positive helicity gluons into
any number of positive helicity gluons.
That is the restriction on polarizations which allows one to obtain the 
explicit formulae for the amplitudes with arbitrary number of legs. Perhaps, 
the simplest procedure to obtain MHV amplitudes has been developed in
\cite{RS1} following ideas of \cite{Ba}, \cite{KS1}.  The procedure
was to obtain the positive helicity gluonic form-factors via a solution
of the self-duality equation and the MHV amplitudes via a solution of
the linearization of YM equations in the self-dual background. In these terms
MHV amplitudes are amplitudes of scattering of two anti-self-dual gluons
into a number of the self-dual ones. (Notice
that a positive helicity gluon is self-dual in the out- state and 
anti-self-dual in the in- state.)

The purpose of the present letter is to point out that the MHV amplitudes
can, after some specifications, be identified as a set of amplitudes
in the spontaneously broken symmetry phase in the Higgs-Yang-Mills (HYM) 
system with adjoint Higgs field in the BPS limit\footnote{We stick on the 
case of SU(2) gauge group, though it is not necessary.}. In the HYM 
interpretation, 
the MHV amplitudes assume more restrictions imposed on the external
particles. First of all, there is a 
kinematical restriction that all the external
particles in this set of amplitudes have zero momentum along a fixed 
coordinate axis. There is also a restriction on polarizations,
namely, all $W^{+}$ bosons have a positive projection of spin onto the fixed
axis, while all   $W^{-}$ bosons have negative spin projection onto it.
We shall call such one-particle states as the BPS ones, because they
are solutions of linearized BPS equations (in $+--$ signature). 
The HYM counterparts of the MHV amplitudes are then the amplitudes
for scattering of a couple of anti-BPS particles into any number
of the BPS ones.

The simplest way to obtain this interpretation is to use the Kaluza-Klein
dimensional reduction (see \cite{bibble}). So, consider 5D YM on 
$M_{4} \times S^{1}$
\begin{equation}
\label{1}
S=\int d^{4}xdy tr \frac{1}{8} F^{2}_{MN}
\end{equation}
with fixed monodromy around $S^{1}$:
\begin{equation}
\label{2}
Pexp\{ \int dyA_{5} \}=exp\{2{\pi}iRa{\sigma}_{3}\}
\end{equation}
where $M,N=0, \ldots , 4$, ${\sigma}_{3}$ - the  Pauli matrix, 
$2{\pi}R$ - circumference of $S^{1}$, $a$ - a constant. 
The low energy effective theory of this
model in the limit $Ra \ll 1$  is 4D HYM system
with adjoint Higgs field having vev $a{\sigma}_{3}$ in the BPS limit.
Let us see how this happens. 

In view of Eq.(\ref{2}), one can choose a gauge so that
\begin{equation}
\label{3}
A=ia{\sigma}_{3}dy+ \left( \ldots \right)
\end{equation}
where $\left( \ldots \right)$ stands for terms periodical in $y$
with periods $2{\pi}R$. Then Eq.(\ref{1}) is rewritten as follows
\begin{equation}
\label{4}
S=\int d^{4}x tr \frac{1}{8} F^{2}_{\mu \nu}+
\frac{1}{4}({\nabla}_{\mu}{\Phi})^{2} +\;KK\;modes
\end{equation} 
where $i{\Phi}\equiv A_{5}$, $\mu, \nu=0, \ldots, 3$. According to
Eq.(\ref{3}), $\Phi$ is assumed to have vev
\begin{equation}
\label{5}
<\Phi>=a{\sigma}_{3}
\end{equation}
The KK modes decouple in the low energy limit. Note that there is no 
Higgs potential in Eq.(\ref{4}) which corresponds to the BPS limit.

The low-energy spectrum of the model Eq.(\ref{4}) with the SU(2) gauge 
group is known to consist
of a ``photon'' (two states), a massless higgs (one state), 
a massive ``$W^{+}$ boson'' (three states) and a massive ``$W^{-}$ bosons''
(three states), both with the mass $4a^{2}$.

One can also describe these states in terms of 5D YM. In order to do this, 
it is convenient to gauge away the monodromy, Eq.(\ref{2}), by the gauge 
transformation with the matrix
\begin{equation}
\label{gauge}
g=e^{-iay\sigma_{3}}
\end{equation}
The price for that will be non-periodicity in $y$ of the gauge
potential $A$:
\begin{equation}
\label{11}
A(y+2{\pi}R)=e^{-2{\pi}iRa\sigma_{3}}A(y)e^{2{\pi}iRa\sigma_{3}}
\end{equation}
One sees that after such gauge transformation the plane waves
corresponding to $W^{\pm}$ bosons get a component of momentum
along y-axis, namely
\begin{eqnarray}
\label{12}
W^{+}:\;k_{5}=-2a\nonumber\\
W^{-}:\;k_{5}=\;\;2a
\end{eqnarray}
Notice that the 5D momenta, made out of the physical 4D ones and 
$k_{5}$ component, are light-like. 

Let us now consider in the 5D YM only amplitudes with the kinematical 
restriction that all the external particles
have zero momentum along one of the axis in the physical 4D space,
say, $k_{3}=0$ for all the external particles. Thus the problem becomes
effectively four-dimensional in the non-physical 4D space spanned by
$x^{0},x^{1},x^{2}, y$. In this non-physical 4D space we shall
use the spinor notation,
\begin{equation}
\label{13}
 x^{\alpha {\dot \alpha}}=\left(\begin{array}{cl}
x^{0}+y      & x^{1}-ix^{2}\\
x^{1}+ix^{2} & x^{0}-y
\end{array} \right) 
\end{equation}
\begin{equation}
\label{14}
 k^{\alpha {\dot \alpha}}=\left(\begin{array}{cl}
k^{0}+k^{5}      & k^{1}-ik^{2}\\
k^{1}+ik^{2} & k^{0}-k^{5}
\end{array} \right) 
\end{equation}
$\alpha, {\dot {\alpha}}=1,2$.
Notice that momenta of the $W$-bosons are light-like in the non-physical
4D space as well as in 5D space. Since they are light-like, they
can be decomposed as follows:
\begin{equation}
\label{15}
 k_{\alpha {\dot \alpha}}=\ae_{\alpha} {\bar \ae}_{\dot \alpha}
\end{equation}
where $\ae_{\alpha}$ and ${\bar \ae}_{\dot \alpha}$ are some spinors.
The reality condition reads
\begin{equation}
\label{16}
 {\bar \ae}_{\dot \alpha}=(\ae_{\alpha})^{\ast}
\end{equation}

Now, consider the MHV amplitudes in this non-physical 4D space, that is,
amplitudes of scattering of two anti-self-dual ``gluons'' into a number
of self-dual ``gluons''. These  amplitudes are represented as follows
\cite{PT},\cite{BG}:
\begin{eqnarray}
\label{17}
A(1^{-},2^{-} \rightarrow 3^{+}, \ldots, d^{+})=
i(\ae_{1},\ae_{2})^{4}
\sum_{\sigma \in S_{d}/Z_{d}}\nonumber\\
\frac{Tr( J_{\sigma(1)} \ldots  J_{\sigma(d)})}
{(\ae_{\sigma(1)},\ae_{\sigma(2)})(\ae_{\sigma(2)},\ae_{\sigma(3)}) 
\ldots (\ae_{\sigma(d)},\ae_{\sigma(1)})}
\end{eqnarray}
where sum is taken over non-cyclically permutations, $J$'s  are the 
color matrixes of the particles,
$\ae$'s are the spinors entering the momenta of the particles, as in 
Eq.(\ref{15}), and the brackets of two $\ae$'s stand for their
contraction with the antisymmetric tensor:
\begin{equation}
\label{18}
(\ae_{1},\ae_{2})\equiv \varepsilon_{\alpha \beta} \ae^{\alpha}\ae^{\beta}
\end{equation}
  
The self-duality (or anti-self-duality) is
assumed in the non-physical 4D space. It is known to be equivalent
to the BPS equation in the HYM terms (in the signature $+--$ in 
the present context).
By studying solutions of linearized field equations and transforming
a self-dual plane wave from the gauge where Eq.(\ref{11}) takes place to
the one where Eq.(\ref{3}) is valid, one can verify that the self-duality in
the non-physical 4D space assumes the restriction on polarizations
of $W^{\pm}$ described above. 

To adopt the amplitudes Eq.(\ref{17}) to the present context one needs
just to specify the color matrixes. The amplitude is nonzero when there are 
equal amounts of $W^{+}$ and $W^{-}$. Then the amplitudes of scattering
of 2 massive anti-BPS particles into $2d-2$ massive BPS particles read
\newpage

\begin{eqnarray}
\label{19}
A(1^{-},2^{-} \rightarrow 3^{+}, \ldots,2d^{+})=
i\frac{1}{d}(\ae_{1},\ae_{2})^{4}
\sum_{\{l\}\{m\}}\nonumber\\
\frac{1}{(\ae_{l_{1}},\ae_{m_{1}})(\ae_{m_{1}},\ae_{l_{2}}) 
\ldots (\ae_{m_{d}},\ae_{l_{1}})}
\end{eqnarray}
where $l_{1},l_{2}, \ldots,l_{d}$ are all different and run over
set numbering $W^{+}$ bosons and  $m_{1},m_{2}, \ldots,m_{d}$
are all different and run over set  numbering  $W^{-}$ bosons .
We would like to emphasize that the spinors entering Eq.(\ref{19}) arose via
decomposition of the type of Eq.(\ref{15}) from the ``non-physical''
momenta of the particles Eq.(\ref{14}). In particular,
the condition Eq.(\ref{12}) in terms of the spinors is rewritten as
\begin{eqnarray}
\label{20}
W^{+}:\; |\ae_{1}|^{2}-|\ae_{2}|^{2} =-4a\nonumber\\
W^{-}:\; |\ae_{1}|^{2}-|\ae_{2}|^{2} =4a
\end{eqnarray}

One can also obtain amplitudes for a different process - scattering of 
two $W$-bosons with zero spin projections onto the fixed axis 
(and zero momentum along the fixed axis) into any number of the BPS 
$W$-bosons. Such amplitudes vanish.

We have explained how the MHV amplitudes in pure YM can be interpreted
as a set of amplitudes of scattering of massive vector bosons in HYM
in the BPS limit (nontrivial vev but no potential for the Higgs field). 
The main tool was the Kaluza-Klein reduction from 5D YM to 4D HYM. 
Analogously, using the Kaluza-Klein reduction from 6D YM to 4D HYM with 
two adjoint Higgs fields, one can obtain from the MHV amplitudes
a set of amplitudes of scattering of massive particles in a theory
with nontrivial Higgs potential. In more complicated versions of
the Kaluza-Klein reduction one can obtain the amplitudes with
fundamental Higgs fields. 

The author benefited a lot from discussions with A.Rosly.
The work was supported in part by the grant INTAS-97-0103.


\begin{thebibliography}{99}
\bibitem{mapa}\ Mangano,M.L., Parke,S., Phys.Rep. {\bf 200}, 301  (1991)
\bibitem{dixon}\ L.Dixon, preprint SLAC-PUB-7106-96, 
 hep-ph/9601359
\bibitem{bern2}\ Z.Bern et al, hep-ph/9706447 
\bibitem{PT} S.Parke and T.Taylor, \Journal{\PRL}{56}{2459}{1986}
\bibitem{B} F.Berends, W.Giele and H.Kuijf, Phys.Lett.B 211 (1988) 91
\bibitem{KS4} K.G.Selivanov, {\em Mod.Phys.Lett.A} 12 {1997} 3087
\bibitem{BG} F.Berends and W.Giele, \Journal{\NPB}{306}{759}{1988}
\bibitem{RS1}  A.A.Rosly and K.G.Selivanov, \Journal{\PLB}{399}{135}{1997}
\bibitem{Ba} W.Bardeen, {\em Prog.Theor.Phys.Suppl.} 123 (1996) 1 
\bibitem{KS1}\ K.G.Selivanov,  ITEP-21-96,  hep-ph/9604206
\bibitem{bibble}\ M.B.Green, J.H.Schwarz and E.Witten,
 Superstring theory, chapter 16, Cambridge University Press, Cambridge, 1988
\end{thebibliography}
\end{document}